\documentclass[aps,prd,twocolumn,showpacs,superscriptaddress,amsmath,graphicx]{revtex4}
\usepackage[dvips]{color,graphicx}
\usepackage{amsmath}
\usepackage{amssymb}
\usepackage{braket}

\newcommand {\vek}[1]{\mathbf{#1}}
\newcommand {\E}[1]{\times 10^{#1}}	
\newcommand {\e}[1]{\mathrm{~#1}}       

\newcommand{\mc}[1]{\mathcal{#1}}

\begin{document}

\title{$b\to d d \bar s$ transition and constraints on new physics in $B^-$ decays}

\author{Svjetlana Fajfer}
\email[Electronic address:]{svjetlana.fajfer@ijs.si}
\affiliation{J. Stefan Institute, Jamova 39, P. O. Box 3000, 1001 Ljubljana, Slovenia}
\affiliation{Department of Physics, University of Ljubljana, Jadranska 19, 1000 Ljubljana, Slovenia}

\author{Jernej Kamenik}
\email[Electronic address:]{jernej.kamenik@ijs.si}
\affiliation{J. Stefan Institute, Jamova 39, P. O. Box 3000, 1001 Ljubljana, Slovenia}

\author{Nejc Ko\v snik}
\email[Electronic address:]{nejc.kosnik@ijs.si}
\affiliation{J. Stefan Institute, Jamova 39, P. O. Box 3000, 1001 Ljubljana, Slovenia}

\date{\today}

\begin{abstract}
  The $b \to dd \bar s$ transition gives extremely
  small branching ratios within the standard model, thus providing an
  appropriate ground for testing new physics.  Using renormalization
  group technique we determine the Wilson coefficients and the mixing
  of the operators which contribute to the $b \to dd \bar s$
  transition. We consider contributions to this decay mode from the
  supersymmetric standard model with and without $\mathcal R$-parity,
  as well as from a model with an additional neutral $Z'$ gauge boson.
  Using Belle and BaBar upper bounds for the $B^- \to \pi^- \pi^- K^+$
  branching ratio we constrain contributions of these new physics
  scenarios.  Then we calculate branching ratios for two- and
  three-body nonleptonic $B^-$ meson decays driven by the $b \to dd \bar
  s$ transition, which might be experimentally accessible.
\end{abstract}


\maketitle

\section{INTRODUCTION}

Among many ongoing searches for physics beyond the standard model
(SM), rare $B$ meson decays seem to offer good opportunities for
discovering new physics.  In particular the experimental results on
decay rates and the parameters describing $CP$ violation in the $B$
meson nonleptonic two-body weak decays such as $B \to \pi K$ and $B
\to \phi K_S$ have attracted a lot of attention during the last few
years~(see e.g.~\cite{Silvestrini:2005zb} and references
  therein).  In the theoretical explanation of these decay rates and
CP violating parameters it is usually assumed that an interplay of the
SM contributions and new physics occurs.  On the other hand, there are
processes of the type $b \to s s \bar d$ and $b \to d d \bar s$ which
are extremely rare within the SM. A careful study of the $b \to s s
\bar d$ transition has been
done~\cite{Huitu:1998vn,Huitu:1998pa,Fajfer:2000ny,Fajfer:2001ht,Fajfer:2000ax}
and the decay $B^- \to \pi^+ K^- K^-$ has been suggested as the most
appropriate mode among possible candidates for experimental searches.
The upper limit was first determined in~\cite{Abbiendi:1999st} and
subsequently constrained by both $B$
factories~\cite{Abe:2002av,Aubert:2003xz}.  These upper bounds gave an
unique opportunity to determine constraints on a variety of scenarios
of new physics such as the minimal supersymmetric standard model
(MSSM) with and without $\mathcal R$-parity violation (RPV),
variations of the two Higgs doublets model (THDM), and models with
additional neutral gauge bosons. Using constraints from this decay
rate the $\Delta S= 2$ two body decays of $B^-$ were
considered~\cite{Fajfer:2000ny} as well as $\Delta S=2$ decays of
$B_c$~\cite{Fajfer:2004fx}.

The $b \to d d \bar s$ transition has not been
subject of such intensive theoretical studies although experimental
information on the upper bound for the $B^- \to \pi^- \pi^- K^+$ decay
rate already exists. Namely, the BaBar collaboration has reported that
${\rm BR} (B^- \to \pi^- \pi^- K^+) < 1.8\E{-6}$~\cite{Aubert:2003xz},
while the Belle collaboration found ${\rm BR} (B^- \to \pi^- \pi^- K^+
)< 4.5\E{-6}$~\cite{Garmash:2003er}.  Hopefully soon the LHC-b would give
even better constraints.

Some time ago Grossman et al.~\cite{Grossman:1999av} have investigated 
the decay mechanisms of $B \to K \pi$ decays and found that new
physics might give important contributions to the relevant
observables.  Within their study of penguin operators which could
receive contributions due to new physics, these authors also included
the effects of the $ \Delta S = -1$ transition.  In their search for
the explanation of the $B \to K \pi$ puzzle, the authors
of~\cite{Barger:2004hn} have investigated the $B \to K \pi$ decay mode
within a model with an extra flavor changing $Z'$ boson, making
predictions for the CP violating asymmetries in these decays. $Z'$
mediated penguin operators have also been considered in many other
scenarios.  Contributions of supersymmetric models with and without
RPV in the same decay channel were discussed
in Ref.~\cite{Arnowitt:2005qz}. The difficulty with this decay mode is that
the SM contribution is the dominant one. The use of quantum
chromodynamics in the treatment of the weak hadronic $B$ meson decays
is not a straightforward procedure.  Numerous theoretical studies have
been attempted to obtain the most appropriate framework to describe
nonleptonic $B$ meson decays to two light meson states.  But even the
most sophisticated approaches such as QCD factorization (BBNS and
SCET)~\cite{Beneke:1999br,Beneke:2000ry,Beneke:2001ev,Beneke:2003zv,Keum:2003js,Bauer:2000ew,Bauer:2000yr,Bauer:2001ct,Bauer:2001yt,Bauer:2004tj,Bauer:2005kd,Williamson:2006hb}
still have parameters which are difficult to obtain from ``first
principles''. Consequently, searches for new physics in decay modes
dominated by SM contributions suffer from large uncertainties.
\par
In this paper we suggest to search for the effects of new physics in
rare decays for which the SM gives negligible contributions.  We only
consider $B^-$ meson decays driven by the $b \to dd \bar s$
transition, since the $\bar u$ anti-quark is a spectator in this
process and one should not worry about possible contributions of the
SM penguins.  The measurement of decay rates for the modes in which
the ``exotic'' $b \to dd \bar s$ transition occurs might give an
unique opportunity to constrain parameters describing new physics. These constraints may then be compared with those obtained from other processes such as $K^0-\bar K^0$ and $B^0-\bar B^0$ transitions.
\par
In the second section, we describe the $b \to d d \bar{s}$ decay and
consider contributions of various new physics models.
First we determine the Wilson coefficients of the hadronic operators
contributing to the effective Hamiltonian in an extended operator
basis, which is applicable for variety of the new physics scenarios.
Namely, we investigate inclusive $b \to dd \bar s$ within the MSSM
with and without RPV, and within an extension of the SM where an
additional flavor changing $Z'$ neutral boson appears.  In Sec. III we
write explicit expressions for the transition matrix elements entering
in exclusive nonleptonic decay rates. Then we consider possible
candidates for the experimental searches.  First we study the
three-body decay $B^- \to \pi^- \pi^- K^+$, which has been already
investigated by both $B$ meson factories.  Then we derive decay rates
for two-body decays $B^- \to \pi^- K^0$, $B^-\to \rho^- K^0$, $B^-\to
\pi^- K^{*0}$, $B^-\to \rho^- K^{*0}$, and three-body decay $B^- \to
\pi^- D^- D_s^+$. In Sec. IV we comment on possibilities to observe
effects of new physics in considered decays and summarize our results.

\section{Inclusive processes}

The effective weak Hamiltonian encompassing the $b \to dd \bar s$
process has been introduced by the authors of~\cite{Grossman:1999av}
in the case of $B \to K \pi$ decays.  Following their notation we
write it as
\begin{equation}
\mathcal H_{\mathrm{eff.}} = \sum_{n=1}^5 \left[ C_n \mathcal O_n + \tilde C_n \tilde {\mathcal O}_n\right],
\end{equation}
where $C_i$ and $\tilde C_i$ denote effective Wilson coefficients
multiplying the complete operator basis of all the four-quark
operators which can contribute to the process $b\to d d \bar s$. We
choose
\begin{eqnarray}
  \mathcal O_1 &=& \bar d^i_L \gamma^{\mu} b^i_L \bar d^j_R \gamma_{\mu} s^j_R, \nonumber\\
  \mathcal O_2 &=& \bar d^i_L \gamma^{\mu} b^j_L \bar d^j_R \gamma_{\mu} s^i_R, \nonumber\\
  \mathcal O_3 &=& \bar d^i_L \gamma^{\mu} b^i_L \bar d^j_L \gamma_{\mu} s^j_L, \nonumber\\
  \mathcal O_4 &=& \bar d^i_R b^i_L \bar d^j_L s^j_R, \nonumber\\
  \mathcal O_5 &=& \bar d^i_R b^j_L \bar d^j_L s^i_R,
\end{eqnarray}
plus additional operators ${\mathcal {\tilde O}_{1,2,3,4,5}}$, with
the chirality exchanges $L\leftrightarrow R$. In these expressions,
the superscripts $i,j$ are $SU(3)$ color indices. All other operators
with the correct Lorentz and color structure can be related to these
by operator identities and Fierz rearrangements. We perform our
calculations of inclusive and exclusive decays at the scale of the $b$
quark mass ($\mu=m_b$), therefore we have to take into account the
renormalization group running of these operators from the interaction
scale $\Lambda$. At leading log order in the strong coupling, the
operators $\mathcal O_{1,2}$ mix with the anomalous dimension matrix
\begin{equation}
\gamma(\mathcal O_1\mathcal O_2) = \frac{\alpha_s}{2\pi}
\left(\begin{array}{cc}
-8 & 0 \\
-3 & 1
\end{array}\right).
\end{equation}
The same holds for operators $\mathcal O_{4,5}$ ($\gamma(\mathcal
O_1\mathcal O_2)=\gamma(\mathcal O_4\mathcal O_5)$) due to Fierz
identities, while the operator $\mathcal O_3$ has anomalous dimension
$\gamma(\mathcal O_3) = \alpha_s/\pi$. Anomalous matrices for chirally
flipped operators $\mathcal{\tilde O}_{1,2,3,4,5}$ are identical to
these.
\par
Within the SM only the operator $\mathcal O_3$ contributes
to the $b \to dd \bar s$ transition at one loop with the Wilson
coefficient
\begin{eqnarray}
  C_3^{SM} &=& \frac{G_F^2}{4 \pi^2} m_W^2 V_{tb} V_{td}^* \Bigg[ V_{ts} V_{td}^* 
  f\left(\frac{m_W^2}{m_t^2}\right) \nonumber\\
  && + V_{cs} V_{cd}^* \frac{m_c^2}{m_W^2} g\left(\frac{m_W^2}{m_t^2},\frac{m_c^2}{m_W^2}\right)\Bigg],
\end{eqnarray}
where the functions $f(x)$ and $g(x,y)$ were given explicitly
in~\cite{Huitu:1998vn}. Using numerical values of the relevant
Cabibbo-Kobayashi-Maskawa (CKM) matrix elements from
PDG~\cite{Eidelman:2004wy} and including the $V_{td}$ phase one finds
$\left|C_3^{SM}\right| \leq 2.5\E{-13}\e{GeV}^{-2}$.  Renormalization
group running from the weak interaction scale to the bottom quark mass
scale, due to anomalous dimension of the operator $\mathcal O_3$,
induces only a small correction factor which can be safely neglected.
The inclusive $b\to d d \bar s$ decay width within the SM is then
\footnote{Our result for SM and MSSM inclusive branching ratio differs
  from the analogue case $b\to s s \bar d$ in Ref.~\cite{Huitu:1998vn}
  by a factor of 4.
}
\begin{equation}
  \Gamma^{SM}_{\mathrm{inc.}} = \frac{\left|C_3^{SM}\right|^2 m_b^5}{48 (2\pi)^3},
\label{eq_Gamma_SM}
\end{equation}
which leads to the branching ratio of the order $10^{-14}$.
\par
Next we discuss contributions of several models containing physics
beyond the SM: the MSSM with and without RPV and a model with an extra
$Z^{\prime}$ boson. For the THDM on the other hand, the contributions
to the $b \to d d \bar s$ transition coming from charged Higgs box
diagrams were found to be negligible. Namely, due to the CKM matrix
elements suppression they would be even smaller than those found
in~\cite{Fajfer:2000ny} for the analogue case of $b\to s s \bar d$.
Consequently we choose to neglect them. In addition, the tree level
neutral Higgs exchange amplitude is proportional to
$|\xi_{db}\xi_{ds}|/m_H^2$, where $\xi_{db}$ and $\xi_{ds}$ are flavor
changing Yukawa couplings and $m_H$ is a common Higgs mass scale. This
ratio is constrained from the neutral meson
mixing~\cite{Huitu:1998pa}. Using presently known values of
$\Delta m_K$ and $\Delta m_B$~\cite{Eidelman:2004wy} one can obtain an
upper bound of $|\xi_{db}\xi_{ds}|/m_H^2<10^{-13}~\mathrm{GeV}^{-2}$
rendering also this contribution negligible~\footnote{In previous work
  on THDM contributions to the complementary $b\to s s \bar d$
  process~\cite{Huitu:1998pa,Fajfer:2000ny,Fajfer:2004fx} the relevant
  effective THDM coupling $|\xi_{sb}\xi_{sd}|/m_H^2$ was not bounded
  from above due to an unknown upper limit on $\Delta m_{B_s}$. With
  the recent two-sided bound on the $B_s$ oscillation frequency from
  the D0 and CDF
  collaborations~\cite{Abazov:2006dm,Gomez-Ceballos:2006gj} it is now
  possible to constrain this contribution to
  $|\xi_{sb}\xi_{sd}|/m_H^2<10^{-12}$. This value is two orders of
  magnitude smaller than the one used in existing studies.
  Correspondingly, all the decay rate predictions for the THDM model
  are diminished by four orders of magnitude and thus rendered
  negligible.}.
\par
In the MSSM, like in the SM, the main contribution comes from the
$\mathcal O_3$ operator, while the corresponding Wilson coefficient is
here
\begin{equation} C_3^{MSSM} = -\frac{\alpha_S^2
    \left(\delta_{21}^d\right)_{LL}^*
    \left(\delta_{13}^d\right)_{LL}}{216 m_{\tilde{d}}^2} [24 x f_6(x)
    + 66 \tilde{f}_6 (x)],\label{eq:Cmssm}
\end{equation} 
as found in analyses~\cite{Gabbiani:1996hi} taking into account only
contributions from the left-handed squarks in the loop.
The recent limits on $\delta_{21}^{d*}
\delta_{13}^d$
~\cite{Khalil:2005qg,Ciuchini:2005kp,Ciuchini:2006dx}
disallow significant contributions from the mixed and the right-handed
squark mass insertion terms. Therefore, we only include the dominant
contributions given in the above expression.  We follow
Ref.~\cite{Ciuchini:2005kp} and take $x=m_{\tilde g^2}/m_{\tilde
  d^2}=1$ and the corresponding values of
$\left|\left(\delta_{13}^d\right)_{LL}(x=1)\right| \leq 0.14$ and
$\left|\left(\delta_{21}^d\right)_{LL}(x=1)\right| \leq 0.042$~\cite{Gabbiani:1996hi}. We
take for the average mass of squarks $m_{\tilde d} = 500\e{GeV}$ and
for the strong coupling constant $\alpha_S=0.12$, and find
$\left|C_3^{MSSM}\right| \leq 1.6\E{-12}\e{GeV}^{-2}$.  Using
Eq.~(\ref{eq_Gamma_SM}) and substituting for the correct Wilson
coefficient one finds the MSSM prediction for the inclusive $b\to d d
\bar s$ decay branching ratio of the order of $10^{-12}$.

If RPV interactions are included in the MSSM, the part of the
superpotential which becomes relevant here is $W =
\lambda^{\prime}_{ijk} L_i Q_j d_k$, where $i,j,$ and $k$ are family
indices, and $L$, $Q$ and $d$ are superfields for the lepton doublet,
the quark doublet, and the down-type quark singlet, respectively. The
tree level effective Hamiltonian receives contributions from the
operators $\mathcal O_{4}$ and $\mathcal{\tilde O}_{4}$ with the
Wilson coefficients defined at the interaction scale $\Lambda \sim
m_{\tilde \nu}$
\begin{eqnarray}
  C_4^{RPV} &=& -\sum_{n=1}^3 \frac{\lambda_{n31}'\lambda_{n12}'^*}{m_{\tilde{\nu}_n}^2}, \nonumber\\ 
  \tilde C_4^{RPV} &=& -\sum_{n=1}^3 \frac{\lambda_{n21}'\lambda_{n13}'^*}{m_{\tilde{\nu}_n}^2}.\label{eq:RPVcouplings} 
\end{eqnarray}
The renormalization group running of the operators induces a common
correction factor for $C_4^{RPV}(\mu) = f_{QCD}(\mu) C_4^{RPV}$ and
$\tilde C_4^{RPV}(\mu) = f_{QCD}(\mu) \tilde C_4^{RPV}$:
\begin{equation}
  f_{QCD}(\mu) = \left\{ 
\begin{array}{ll}
  \left[\frac{\alpha_s(\mu)}{\alpha_s(\Lambda)}\right]^{24/23}, & \Lambda < m_t \\
  \left[\frac{\alpha_s(\mu)}{\alpha_s(m_t)}\right]^{24/23}\left[\frac{\alpha_s(m_t)}{\alpha_s(\Lambda)}\right]^{24/21}, & \Lambda > m_t \\
\end{array}\right\}, 
\end{equation}
which evaluates to $f_{QCD}(m_b)\simeq 2$ for a range of sneutrino
masses between $100~\mathrm{GeV}\lesssim m_{\tilde \nu} \lesssim
1~\mathrm{TeV}$. 
In addition, the mixing with the operators $\mathcal O_5$ and
$\tilde{\mathcal O}_5$ induces a small contribution to the Wilson
coefficients $C_5^{RPV}(\mu) = \tilde f_{QCD}(\mu) C_4^{RPV}$ and
$\tilde C_5^{RPV}(\mu) = \tilde f_{QCD}(\mu) \tilde C_4^{RPV}$:
\begin{widetext}
\begin{equation}
\tilde f_{QCD}(\mu) =  \frac{1}{3}\left\{ 
  \begin{array}{ll}
    \left[\frac{\alpha_s(\mu)}{\alpha_s(\Lambda)}\right]^{24/23}-\left[\frac{\alpha_s(\mu)}{\alpha_s(\Lambda)}\right]^{-3/23}, & \Lambda < m_t \\
    \left[\frac{\alpha_s(\mu)}{\alpha_s(m_t)}\right]^{24/23}\left[\frac{\alpha_s(m_t)}{\alpha_s(\Lambda)}\right]^{24/21}-\left[\frac{\alpha_s(\mu)}{\alpha_s(m_t)}\right]^{-3/23}\left[\frac{\alpha_s(m_t)}{\alpha_s(\Lambda)}\right]^{-3/21}, & \Lambda > m_t \\
  \end{array}\right\} 
\end{equation}
\end{widetext}
which is of the order $\tilde f_{QCD}(m_b) \simeq 0.4$ for the chosen
sneutrino mass range. The relevant part of the effective Hamiltonian
we use in this scenario is then 
\begin{align}
  \mathcal H_{\mathrm{eff.}}^{RPV} =& f_{QCD}(\mu) \left[C_4^{RPV}
    \mathcal O_4(\mu) + \tilde C_4^{RPV} \tilde {\mathcal O}_4(\mu)
  \right]\nonumber\\
  & +\tilde{f}_{QCD}(\mu) \left[C_4^{RPV} \mathcal O_5(\mu) + \tilde
    C_4^{RPV} \tilde {\mathcal O}_5(\mu) \right].
\end{align}
We neglect the $\tilde f_{QCD}$ suppressed contributions of $\mc O_5$,
$\tilde{\mc O}_5$ to the amplitudes in the cases where the operators
$\mc O_4$, $\tilde{\mc O}_4$ give non-zero
contribution.\label{suppressedO5} The inclusive $b\to d d \bar s$
decay rate induced by the RPV model becomes
\begin{equation}
  \Gamma^{RPV}_{\mathrm{inc.}} = \frac{m_b^5 f^2_{QCD}(m_b)}{256 (2\pi)^3}\left(| C_4^{RPV}|^2 + | \tilde C_4^{RPV}|^2\right).
\label{eq_Gamma_RPV}
\end{equation}
Present experimental bounds on the individual RPV couplings
contributing to the effective Wilson coefficients $C_4^{RPV}$ and
$\tilde C_4^{RPV}$ do not constrain this mode, and we extract the
bounds on the relevant combination from exclusive decays in
Sec.~\ref{discussion}.

In many extensions of the SM~\cite{Langacker:2000ju} an additional
neutral gauge boson appears. Heavy neutral bosons are also present in
grand unified theories, superstring theories and theories with large
extra dimensions~\cite{Erler:1999nx}.  This induces contributions to
the effective tree level Hamiltonian from the operators $\mathcal
O_{1,3}$ as well as $\mathcal {\tilde O}_{1,3}$.  Following
\cite{Langacker:2000ju,Erler:1999nx}, the Wilson coefficients for the
corresponding operators read at the interaction scale $\Lambda \sim
m_{Z'}$
\begin{equation} \label{eq:zPrimeOperators}
\begin{array}{rclrcl}
  C_1^{Z'} &=& -\frac{4G_F y}{\sqrt{2}} B_{12}^{d_L} B_{13}^{d_R}, & \tilde C_1^{Z'} &=& -\frac{4G_F y}{\sqrt{2}} B_{12}^{d_R} B_{13}^{d_L},\\
  C_3^{Z'} &=& -\frac{4G_F y}{\sqrt{2}} B_{12}^{d_L} B_{13}^{d_L}, & \tilde C_3^{Z'} &=& -\frac{4G_F y}{\sqrt{2}} B_{12}^{d_R} B_{13}^{d_R}, 
\end{array}
\end{equation}
where $y = (g_2/g_1)^2 (\rho_1 \sin^2 \theta+ \rho_2 \cos^2 \theta)$
and $\rho_i = m_W^2/m_i^2 \cos^2\theta_W$. In this expression $g_1$,
$g_2$, $m_1$ and $m_2$ stand for the gauge couplings and masses of the
$Z$ and $Z'$ bosons, respectively, while $\theta$ is their mixing
angle.  Again renormalization group running induces corrections and
mixing between the operators.  As already mentioned, the mixing of
operators $\mathcal O_{1,2}$ and their chirally flipped counterparts
is identical to that of operators $\mathcal O_{4,5}$ since these
operators are connected via Fierz rearrangement. Thus the same scaling
and mixing factors $f_{QCD}$ and $\tilde f_{QCD}$ apply. For the
operator $\mathcal O_3$ on the other hand the renormalization can be
written as $C^{Z'}_3(\mu) = f'_{QCD}(\mu) C^{Z'}_3$ with
\begin{equation}
  f'_{QCD}(\mu) = \left\{ 
    \begin{array}{ll}
      \left[\frac{\alpha_s(\mu)}{\alpha_s(\Lambda)}\right]^{-6/23}, & \Lambda < m_t \\
      \left[\frac{\alpha_s(\mu)}{\alpha_s(m_t)}\right]^{-6/23}\left[\frac{\alpha_s(m_t)}{\alpha_s(\Lambda)}\right]^{-6/21}, & \Lambda > m_t \\
    \end{array}\right\}. 
\end{equation}
In particular for a common $Z'$ boson scale of $m_{Z'}\simeq
500~\mathrm{GeV}$~\cite{Langacker:2000ju} one gets numerically
$f_{QCD} (m_b) \simeq 2$, $\tilde f_{QCD} (m_b) \simeq 0.4 $ and
$f'_{QCD} (m_b) \simeq 0.8$. The full contributing part of the
effective Hamiltonian in this case is
\begin{eqnarray}
  \mathcal H_{\mathrm{eff.}}^{Z'} &=& f_{QCD}(\mu) \left[C_1^{Z'} \mathcal O_1(\mu) + \tilde C_1^{Z'} \tilde {\mathcal O}_1(\mu) \right] \nonumber\\
  &&+ \tilde f_{QCD}(\mu) \left[C_1^{Z'} \mathcal O_2(\mu) + \tilde C_1^{Z'} \tilde {\mathcal O}_2(\mu) \right] \nonumber\\
  &&+ f'_{QCD}(\mu) \left[C_3^{Z'} \mathcal O_3(\mu) + \tilde C_3^{Z'} \tilde {\mathcal O}_3(\mu) \right].\label{eq:zHamiltonian}
\end{eqnarray}
For the inclusive $b\to d d \bar s$ decay rate the $\mathcal O_2$ and
$\tilde {\mathcal O_2} $ are numerically suppressed due to the $\tilde
f_{QCD}$ factor and we write
\begin{eqnarray}
  \Gamma^{Z'}_{\mathrm{inc.}} &=& \frac{m_b^5}{192 (2\pi)^3}\Big[ 3 f^2_{QCD}(m_b) \left(|C^{Z'}_1|^2+|\tilde C^{Z'}_1|^2\right) \nonumber\\
  && + 4 f'^2_{QCD}(m_b)\left(|C^{Z'}_3|^2+|\tilde C^{Z'}_3|^2\right)\Big].
  \label{eq_Gamma_Z}
\end{eqnarray}
In Sec.~\ref{discussion} we discuss bounds on Wilson coefficients
$C_{1,3}^{Z'}$ and $\tilde{C}_{1,3}^{Z'}$ which might be estimated
from the $B^- \to \pi^- \pi^- K^+$ decay rate.


\section{Exclusive $B^-$ decay modes}
In calculating decay rates of various $B$ meson decay modes based on
the $b \to dd\bar{s}$ quark transition, one has to calculate matrix
elements of the effective Hamiltonian operators between meson states.
As a first approximation, we use the naive factorization of three-body
amplitudes, and express the resulting two-body transition amplitudes
between mesons in terms of the standard weak transition form factors
(\ref{eq:PtoP},\ref{eq:PtoV}), as dictated by the Lorentz covariance.
For the $B\to \pi (\rho)$ transitions we use form factors calculated
in the relativistic constituent quark model, with numerical input from
the lattice QCD at high momentum transfer
squared~\cite{Melikhov:2000yu}. For the $D_s \to D$ and $K \to \pi$
transition form factors we use results of Refs.~\cite{Fajfer:2004fx,
  Fajfer:1999hh} where heavy meson effective theory and chiral
Lagrangian approach were used.
 
For the decays of the $B^-$ meson to three pseudoscalar mesons $P$,
$P_1$ and $P_2$ we first derive a general expression for the
factorized matrix element of the $\mathcal O_3$ operator, relevant in
the framework of SM~(MSSM)
\begin{widetext}
\begin{align}
  \Braket{P_2 (p_2) P_1 (p_1) | \bar{d} \gamma_\mu s |0} &\Braket{P(p)
    | \bar{d} \gamma^\mu b| B^-(p_B) }
  = (t-u) F_1^{P_2 P_1} (s) F_1^{PB} (s)\nonumber\\
  &+ \frac{(m_{P_1}^2-m_{P_2}^2)(m_B^2-m_P^2)}{s} \left[ F_1^{P_2 P_1}
    (s) F_1^{PB} (s) -F_0^{P_2 P_1} (s) F_0^{PB} (s) \right].
  \label{eq:3bodySM}
\end{align}
\end{widetext}
Because only vector currents contribute in the above expression, it
also applies for operators $\tilde {\mc O}_3$, $\mc O_1$ and $\tilde
{\mc O}_1$.  Form factors $F_1$ and $F_0$ are defined in the
Appendix~A and the Mandelstam kinematical variables are $s=(p_B-p)^2$,
$t=(p_B-p_1)^2$ and $u=(p_B-p_2)^2$.

In the context of RPV, the contributions of the operators $\mc{O}_4$
and $\tilde {\mc O}_4$ to hadronic amplitudes are dominant.  One can
use the Dirac equation to express scalar~(pseudoscalar) density
operators in terms of derivatives of vector~(axial-vector) currents
\begin{subequations} \label{eq:diracTrick}
  \begin{align}
    \bar{q}_i q_j &= \frac{i \partial_\mu(\bar{q}_i \gamma^{\mu} q_j)}{m_{q_j} - m_{q_i}},\\
    \bar{q}_i \gamma^5 q_j &= -\frac{i \partial_\mu (\bar{q}_i
      \gamma^{\mu} \gamma^5 q_j)}{m_{q_j} + m_{q_i}}.
  \end{align}
\end{subequations}
Using these relations we derive an expression for the factorized matrix
element of the $\mc O_4$ and $\tilde{\mc O}_4$ operators, contributing only 
with their scalar parts
\begin{multline}
  \Braket{P_2 (p_2) P_1 (p_1) | \bar{d} s |0} \Braket{P(p) | \bar{d} b | B^-(p_B) } = \\
  \frac{(m_{P_1}^2-m_{P_2}^2)(m_B^2-m_P^2)}{(m_b - m_d)(m_s - m_d)}
  F_0^{P_2 P_1} (s) F_0^{PB} (s) \label{eq:3bodyO4}.
\end{multline}

In the case of the $Z'$ model one encounters contributions of the
operators $\mc O_{1,2,3}$ and $\tilde{\mc O}_{1,2,3}$.  The color
non-singlet operators $\mc O_2$ and $\tilde {\mc O}_2$ can be Fierz
rearranged to $\mc O_4$ and $\tilde {\mc O}_4$ and then
Eq.~(\ref{eq:3bodyO4}) applies as well.  Remaining operators are all
of the $V\pm A$ form and their contribution to the amplitude is
already given by Eq.~(\ref{eq:3bodySM}).

In two-body decays with a vector meson $V$ and a pseudoscalar meson
$P$ in the final state we sum over the polarizations of $V$. The sum
in our case reduces to
\begin{equation}
  \label{eq:polSum} \sum_{\epsilon_V} \left|\epsilon_V^*(p_V)\cdot p_B \right|^2
  = \frac{\lambda(m_B^2,m_V^2,m_P^2)}{4m_V^2}, 
\end{equation} 
where $\epsilon_V$ is the polarization vector of $V$ and $\lambda$ is
defined as $\lambda(x,y,z) = (x+y+z)^2-4(xy+yz+zx)$. 

For decay to two vector mesons in the final state we use the helicity
amplitudes formalism as described in Ref.~\cite{Kramer:1991xw}.
Non-polarized decay rate is expressed as an incoherent sum of helicity
amplitudes
\begin{equation}
  \Gamma=\frac{|\vek{p}_1|}{8 \pi m_B^2} \left(\left|H_0\right|^2+\left|H_{+1}\right|^2+\left|H_{-1}\right|^2 \right),
\end{equation}
where $\vek{p}_1$ is momentum of the vector meson in $B^-$ meson rest
frame and helicity amplitudes are expressed as
\begin{subequations}
\begin{align}
  H_{\pm 1} &= a \pm \frac{\sqrt{\lambda(m_B^2,m_1^2,m_2^2)}}{2 m_1 m_2} c,\\
  H_0 &= -\frac{m^2-m_1^2-m_2^2}{2m_1 m_2} a
  -\frac{\lambda(m_B^2,m_1^2,m_2^2)}{4 m_1^2 m_2^2}b.
\end{align}
\end{subequations}
Vector meson masses are denoted by $m_{1,2}$, while definition of the
constants $a$, $b$ and $c$ is given by general Lorentz decomposition
of the polarized amplitude
\begin{align} \label{eq:helicityDecomposition}
  H_\lambda=\epsilon_{1\mu}^*(\lambda) \epsilon_{2\nu}^*(\lambda) &\left(a g^{\mu\nu} + \frac{b}{m_1 m_2}p_B^\mu p_B^\nu\right.\nonumber\\
  &+ \left.\frac{ic}{m_1 m_2} \epsilon^{\mu\nu\alpha\beta}p_{1\alpha}
    p_{2\beta} \right),
\end{align}
where $\epsilon_{1,2}$ and $p_{1,2}$ are the vector mesons polarizations and momenta.

\subsection{$B^- \to \pi^- \pi^- K^+$}
Hadronic matrix element entering in the amplitude for $B^- \to \pi^-
\pi^- K^+$ in SM~(MSSM) is readily given by Eq.~(\ref{eq:3bodySM})
after identifying $P = \pi^-$, $P_1 = K^+$, $P_2 = \pi^-$ and using
appropriate form factors given in the Appendix~A.
Eq.~(\ref{eq:3bodyO4}) is used instead for RPV, while the $Z'$
amplitude incorporates both Eqs.~(\ref{eq:3bodySM}) and
(\ref{eq:3bodyO4}).  There are two contributions in each model to this
mode, with an additional term with the $u \leftrightarrow s$
replacement in Eqs.  (\ref{eq:3bodySM}) and (\ref{eq:3bodyO4}),
representing an interchange of the two pions in the final state.
After phase space integration, the decay rates can be written very
compactly with only Wilson coefficients left in symbolic form:
\begin{align}
  &\Gamma^{(MS)SM}_{\pi\pi K} = \left|C_3^{(MS)SM}\right|^2 \times
  2.0\E{-3}\e{GeV}^5,\\
  &\Gamma^{RPV}_{\pi\pi K} = \left|C_4^{RPV}+\tilde C_4^{RPV
    }\right|^2 \times
  9.2\E{-3}\e{GeV}^5,\label{eq:3bodyRPVrate}\\
  &\Gamma^{Z'}_{\pi\pi K} = \left|C_1^{Z'} + \tilde C_1^{Z'}\right|^2 \times 1.0\E{-2}\e{GeV}^5\nonumber\\
  &\phantom{\Gamma^{Z'}_{\pi\pi K} =}+\left|C_3^{Z'} + \tilde C_3^{Z'}\right|^2 \times 1.3\E{-3}\e{GeV}^5\nonumber\\
  &\phantom{\Gamma^{Z'}_{\pi\pi K} =}+\mathrm{Re}\left[\left(C_1^{Z'} + \tilde C_1^{Z'}\right) \left(C_3^{Z'} + \tilde C_3^{Z'}\right)^*\right]\nonumber\\
  &\phantom{\Gamma^{Z'}_{\pi\pi K} =+}\times 6.7\E{-3}\e{GeV}^5
  \label{eq:3bodyGammaZ}.
\end{align}

\subsection{$B^- \to \pi^- D^- D_s^+$}
In calculation of the $B^- \to \pi^- D^- D_s^+$ decay rate again we
use Eqs.~(\ref{eq:3bodySM}) and (\ref{eq:3bodyO4}) now with
substitutions $P=\pi^-$, $P_1 = D_s^+$ and $P_2 = D^-$. Numerically
this yields
\begin{align}
  &\Gamma^{(MS)SM}_{\pi D D_s} = \left|C_3^{(MS)SM}\right|^2 \times
  8.7\E{-9}\e{GeV}^5,\\
  &\Gamma^{RPV}_{\pi D D_s} = \left|C_4^{RPV}+\tilde C_4^{RPV
    }\right|^2 \times 8.4\E{-5}\e{GeV}^5,\\
  &\Gamma^{Z'}_{\pi D D_s} = \left|C_1^{Z'} + \tilde C_1^{Z'}\right|^2 \times 1.5\E{-5}\e{GeV}^5\nonumber\\
  &\phantom{\Gamma^{Z'}_{\pi D D_s} =}+\left|C_3^{Z'} + \tilde C_3^{Z'}\right|^2 \times 5.5\E{-9}\e{GeV}^5\nonumber\\
  &\phantom{\Gamma^{Z'}_{\pi D D_s} =}+\mathrm{Re}\left[\left(C_1^{Z'} + \tilde C_1^{Z'}\right) \left(C_3^{Z'} + \tilde C_3^{Z'}\right)^*\right]\nonumber\\
  &\phantom{\Gamma^{Z'}_{\pi D D_s} =+}\times 5.7\E{-7}\e{GeV}^5.
\end{align}
These decay rates are suppressed due to the small phase space in
comparison to the rates of the $B^- \to \pi^- \pi^- K^+$ decay.

\subsection{$B^- \to \pi^- K^0$}
The authors of Ref.~\cite{Grossman:1999av} addressed this decay mode
as the wrong kaon mode, being highly suppressed in the SM compared to
the decay with $\bar{K}^0$ in the final state. The operators $\mc
O_{1,3}$ and $\tilde{\mc O}_{1,3}$ that are present in SM~(MSSM) and
$Z'$ model have the following contribution:
\begin{align}
  &\Braket{K^0(p_K) | \bar{d} \gamma_\mu \gamma^5 s | 0}
  \Braket{\pi^-(p_\pi) | \bar{d}\gamma^\mu b |B^- (p_B)}
  \nonumber\\
  &\quad= i (m_B^2-m_\pi^2) f_K F_0^{\pi B} (m_K^2).
\end{align}
Operators $\mc O_4$ and $\mc{\tilde O}_4$, relevant for the RPV and
$Z'$ models result in
\begin{align}
&\Braket{K^0(p_K) | \bar{d} \gamma^5 s | 0} \Braket{\pi^-(p_\pi) | \bar{d} b | B^-(p_B)}  \nonumber\\
  &\quad=\frac{i m_K^2 (m_B^2 - m_\pi^2)}{(m_b-m_d)(m_s+m_d)} f_K F_0^{\pi B} (m_K^2).
\end{align}
However, in the latter two models, the two
chirally flipped contributions to the amplitude have opposite signs,
resulting in a slightly different combination of Wilson coefficients
in comparison with the $B^- \to \pi^- \pi^- K^+$ decay rate
\begin{align}
  &\Gamma^{(MS)SM}_{\pi K} = \left|C_3^{(MS)SM}\right|^2 \times
  3.9\E{-4}\e{GeV}^5,\\
  &\Gamma^{RPV}_{\pi K} = \left|C_4^{RPV}-\tilde C_4^{RPV
    }\right|^2\times
  4.9\E{-4}\e{GeV}^5\label{eq:RPVratepiK},\\
  &\Gamma^{Z'}_{\pi K} = \left|C_1^{Z'} - \tilde C_1^{Z'}\right|^2 \times 9.5\E{-4}\e{GeV}^5 \nonumber\\
  &\phantom{\Gamma^{Z'}_{\pi K} =}+\left|C_3^{Z'} - \tilde C_3^{Z'}\right|^2 \times 2.5\E{-4}\e{GeV}^5 \nonumber\\
  &\phantom{\Gamma^{Z'}_{\pi K} =}-\mathrm{Re}\left[\left(C_1^{Z'} - \tilde C_1^{Z'}\right)\left(C_3^{Z'} - \tilde C_3^{Z'}\right)^*\right]\nonumber\\
  &\phantom{\Gamma^{Z'}_{\pi K} =-}\times
  9.8\E{-4}\e{GeV}^5.\label{eq:ZratepiK}
\end{align}

\subsection{$B^- \to \rho^- K^0$}
Using form factors parameterization (\ref{eq:PtoV}) of the
pseudoscalar to vector meson transition we derive the following two
factorized matrix elements of axial-vector and pseudoscalar operators:
\begin{align}
  &\Braket{K^0(p_K) | \bar{d} \gamma_\mu \gamma^5 s | 0}
  \Braket{\rho^-(\epsilon_\rho, p_\rho) |\bar{d}\gamma^\mu \gamma^5 b
    |B^- (p_B)}
  \nonumber\\
  &\quad= -2 m_\rho  f_K A^{\rho B}_0(m_K^2) \epsilon_\rho^* \cdot p_B,\\
  &\Braket{K^0(p_K) | \bar{d} \gamma^5 s |
    0}\Braket{\rho^-(\epsilon_\rho, p_\rho) |\bar{d} \gamma^5 b |B^-
    (p_B)}
  \nonumber\\
  &\quad= \frac{2m_\rho m_K^2}{(m_b+m_d)(m_s+m_d)} f_K A^{\rho
    B}_0(m_K^2)\epsilon_\rho^*\cdot p_B.
\end{align}
Finally, we sum over polarizations of the $\rho$ meson using
Eq.~(\ref{eq:polSum}), and the unpolarized decay rates read
\begin{align}
  &\Gamma^{(MS)SM}_{\rho K} = \left|C_3^{(MS)SM}\right|^2 \times 3.9\E{-4}\e{GeV}^5,\\
  &\Gamma^{RPV}_{\rho K} = \left|C_4^{RPV} + \tilde C_4^{RPV} \right|^2\times 4.9\E{-4}\e{GeV}^5,\\
  &\Gamma^{Z'}_{\rho K} = \left|C_1^{Z'} + \tilde C_1^{Z'}\right|^2\times 2.3\E{-3}\e{GeV}^5\nonumber\\
 &\phantom{\Gamma^{Z'}_{\rho K} =}+\left|C_3^{Z'} + \tilde C_3^{Z'}\right|^2\times 2.5\E{-4}\e{GeV}^5\nonumber\\
  &\phantom{\Gamma^{Z'}_{\rho K} =}-\mathrm{Re}\left[\left(C_1^{Z'} + \tilde C_1^{Z'}\right)\left(C_3^{Z'} + \tilde C_3^{Z'}\right)^*\right]\nonumber\\
  &\phantom{\Gamma^{Z'}_{\rho K} =-}\times 1.5\E{-3}\e{GeV}^5.
\end{align}

\subsection{$B^- \to \pi^- K^{*0}$}
Factorized matrix element is here a product of vector meson $K^{*0}$
creation amplitude~(\ref{eq:KstardecayConstant}) and $B^- \to \pi^-$
transition amplitude. Operators which involve vector currents result in
\begin{align}
  \Braket{K^{*0}(\epsilon_K,p_K) | \bar{d} \gamma_\mu s | 0} &\Braket{\pi^-(p_\pi) | \bar{d} \gamma^\mu b|B^-(p_B)} = \nonumber\\
& 2 g_{K^*} F_1^{\pi B}(m_{K^*}^2) \epsilon_K^*\cdot p_B,
\end{align}
while the density operators $\mc O_4$ and $\tilde{\mc O}_4$ do not
contribute, as a result of Eqs.~(\ref{eq:diracTrick}) and
(\ref{eq:KstardecayConstant}). Consequently, in the RPV model this
mode is dominated by the operators $\mc O_5$ and $\tilde{\mc O}_5$
which are, as mentioned in Sec.~\ref{suppressedO5}, suppressed by
the renormalization group running. Using Fierz rearrangements, we
write them down as $\mc O_1$, $\tilde{\mc O}_1$ and yield an additional
$1/2$ suppression factor.
\begin{align}
  &\Gamma^{(MS)SM}_{\pi K^*} = \left|C_3^{(MS)SM}\right|^2 \times
  7.4\E{-4}\e{GeV}^5,\\
  &\Gamma^{RPV}_{\pi K^*} = \left|C_4^{RPV} + \tilde C_4^{RPV}\right|^2\times 2.9\E{-5}\e{GeV}^5\\
  &\Gamma^{Z'}_{\pi K^*} = \left|C_1^{Z'} + \tilde C_1^{Z'}\right|^2\times 2.9\E{-3}\e{GeV}^5\nonumber\\
  &\phantom{\Gamma^{Z'}_{\pi K^*} =}+\left|C_3^{Z'} + \tilde C_3^{Z'}\right|^2\times 4.7\E{-4}\e{GeV}^5\nonumber\\
  &\phantom{\Gamma^{Z'}_{\pi K^*} =}+\mathrm{Re}\left[\left(C_1^{Z'} + \tilde C_1^{Z'}\right)\left(C_3^{Z'} + \tilde C_3^{Z'}\right)^*\right]\nonumber\\
  &\phantom{\Gamma^{Z'}_{\pi K^*} =-}\times 2.3\E{-3}\e{GeV}^5.
\end{align}

\subsection{$B^- \to \rho^- K^{*0}$}
Like in the previous case, this mode only receives contributions from
the renormalization group suppressed RPV terms. We calculate
unpolarized hadronic amplitudes of the operators $\mc O_{1,3}$ and
$\tilde{\mc O}_{1,3}$ by utilizing the helicity amplitudes formalism.
Using form factor decomposition~(\ref{eq:KstardecayConstant},
\ref{eq:PtoV}), we write down the expression for the polarized
amplitude~(\ref{eq:helicityDecomposition}) and identify
constants $a$, $b$ and $c$:
\begin{subequations}
\begin{align}
a&=-\frac{i}{4}(m_B+m_\rho)g_{K^*} A_1^{\rho B} (m_{K^*}^2)(C-\tilde C),\\
b&=\frac{i}{2} \frac{m_{K^*}m_\rho}{m_B+m_\rho}g_{K^*} A_2^{\rho B}(m_{K^*}^2)(C -\tilde C),\\
c&=-\frac{i}{2} \frac{m_{K^*}m_\rho}{m_B+m_\rho}g_{K^*} V^{\rho B}(m_{K^*}^2)(C +\tilde C).
\end{align}
\end{subequations}
$C$ and $\tilde C$ are combinations of the Wilson coefficients present
in a considered model. We have $C=C_3^{(MS)SM}$, $\tilde C=0$ in the
SM~(MSSM), $C=-\tilde{f}_{QCD}(m_b) C_4^{RPV}/2$, $\tilde
C=-\tilde{f}_{QCD}(m_b) \tilde C_4^{RPV}/2$ in the case of the RPV
model and $C=f_{QCD}(m_b) C_1^{Z'} + f_{QCD}'(m_b) C_3^{Z'}$, $\tilde
C = f_{QCD}(m_b) \tilde C_1^{Z'} + f_{QCD}'(m_b) \tilde C_3^{Z'}$ in
the $Z'$ model. Decay rates are then
\begin{align}
  &\Gamma^{(MS)SM}_{\rho K^*} = \left|C_3^{(MS)SM}\right|^2 \times
  9.2\E{-4}\e{GeV}^5,\\
  &\Gamma^{RPV}_{\rho K^*} = \left|C_4^{RPV} + \tilde C_4^{RPV}\right|^2\times 1.4\E{-6}\e{GeV}^5\nonumber\\
  &\phantom{\Gamma^{RPV}_{\rho K^*} =}+\left|C_4^{RPV} - \tilde C_4^{RPV}\right|^2\times 3.5\E{-5}\e{GeV}^5,\\
  &\Gamma^{Z'}_{\rho K^*} = \left|C_1^{Z'} + \tilde C_1^{Z'}\right|^2\times 1.4\E{-4}\e{GeV}^5\nonumber\\
  &\phantom{\Gamma^{Z'}_{\rho K^*} =}+\left|C_1^{Z'} - \tilde C_1^{Z'}\right|^2\times 3.5\E{-3}\e{GeV}^5\nonumber\\
  &\phantom{\Gamma^{Z'}_{\rho K^*} =}+\left|C_3^{Z'} + \tilde C_3^{Z'}\right|^2\times 2.2\E{-5}\e{GeV}^5\nonumber\\
  &\phantom{\Gamma^{Z'}_{\rho K^*} =}+\left|C_3^{Z'} - \tilde C_3^{Z'}\right|^2\times 5.7\E{-4}\e{GeV}^5\nonumber\\
  &\phantom{\Gamma^{Z'}_{\rho K^*} =}+\mathrm{Re}\left[\left(C_1^{Z'} + \tilde C_1^{Z'}\right)\left(C_3^{Z'} + \tilde C_3^{Z'}\right)^*\right]\nonumber\\
  &\phantom{\Gamma^{Z'}_{\rho K^*} =+}\times 1.1\E{-4}\e{GeV}^5\nonumber\\
  &\phantom{\Gamma^{Z'}_{\rho K^*} =}+\mathrm{Re}\left[\left(C_1^{Z'} - \tilde C_1^{Z'}\right)\left(C_3^{Z'} - \tilde C_3^{Z'}\right)^*\right]\nonumber\\
  &\phantom{\Gamma^{Z'}_{\rho K^*} =+}\times 2.8\E{-3}\e{GeV}^5 .
\end{align}

\section{DISCUSSION AND RESULTS}\label{discussion}

We have investigated the $b \to d d \bar s$ transition within the SM,
MSSM without and with RPV terms and within a model with an extra $Z'$
gauge boson. The SM contribution leads to extremely small branching
ratio for this transition.

First we have calculated the effects of QCD on the Wilson coefficients
caused by the renormalization group running. The moderate increase of
the MSSM compared to the SM predictions is still too insignificant for
any experimental search. The MSSM with RPV terms, however, might give
significant contributions and a possibility to shrink down the
parameter space even further.  The $Z'$ model exhibits its structure
through interplay of different interaction scale couplings and might
also give opportunity to constrain its relevant parameters.  
In the case of the two Higgs doublet
model we do not expect any sizable effect as already noticed in the
case of $b \to s s \bar d$ decays~\cite{Fajfer:2000ny}.

In the $b \to d d \bar s$ decay a particular combination of the model
parameters appear which can be constrained using the $B^- \to \pi^-
\pi^- K^+$ decay mode. In our calculation we have relied on the naive
factorization approximation, which is as a first approximation
sufficient to obtain correct gross features of new physics effects.
One might think that the nonfactorizable contributions might induce
large additional uncertainties, but we do not expect them to change
the order of magnitude of our predictions.  Additional uncertainties
might originate in the poor knowledge of the input parameters such as
form factors. However, we do not expect these to exceed more than
$30\%$.
 
Using the stringest experimental bound for the $\mathrm{BR}(B^- \to
\pi^- \pi^- K^+) < 1.8 \times 10^{-6}$ and normalizing the masses of
sneutrinos to a common mass scale of $100\e{GeV}$ we derive bounds on
the RPV terms given in Eq.~(\ref{eq:RPVcouplings})
\begin{equation}
  \left| \sum_{n=1}^3 \left(\frac{100~\mathrm{GeV}}{m_{\tilde\nu_n}}\right)^2 \left(\lambda_{n31}' \lambda_{n12}'^* + \lambda_{n21}' \lambda_{n13}'^*\right)\right| < 8.9\E{-5}.
\label{par-R}
\end{equation}
Complementary bounds coming from measurements of $K^0-\bar{K^0}$ and $B^0-\bar{B^0}$ 
mixings have been established in Refs.~\cite{Kundu:2004cv,Chemtob:2004xr}
\begin{subequations}
\begin{align}
&\left|\mathrm{Re}\left[\sum_{n=1}^3
\left(\frac{100\e{GeV}}{m_{\tilde\nu_n}}\right)^2\lambda_{n31}'
\lambda_{n12}'^* \right]\right|<2.6\E{-6},\label{eq:KKbarre}\\ 
&\left|\mathrm{Im}\left[\sum_{n=1}^3
\left(\frac{100\e{GeV}}{m_{\tilde\nu_n}}\right)^2\lambda_{n31}'
\lambda_{n12}'^* \right]\right|<2.9\E{-8},\label{eq:KKbarim}\\
&\left|\mathrm{Re}\left[\sum_{n=1}^3
\left(\frac{100\e{GeV}}{m_{\tilde\nu_n}}\right)^2\lambda_{n21}'
\lambda_{n13}'^* \right]\right|<2.9\E{-4}.\label{eq:BBbarre}
\end{align}
\end{subequations}
From Eqs.~(\ref{eq:KKbarre}) and (\ref{eq:KKbarim}) it becomes apparent that
the $\lambda_{n31}' \lambda_{n12}'^*$ term is negligible in
Eq.~(\ref{par-R}), and the bound becomes simpler
\begin{equation}
  \left| \sum_{n=1}^3 \left(\frac{100~\mathrm{GeV}}{m_{\tilde\nu_n}}\right)^2 \lambda_{n21}' \lambda_{n13}'^*\right| < 8.9\E{-5},
\end{equation}
now being more restrictive than Eq.~(\ref{eq:BBbarre}), obtained from
$B^0-\bar{B^0}$ mixing.

Assuming that new physics arises due to an extra $Z'$ gauge boson we
derive bounds on the parameters given in
Eq.~(\ref{eq:zPrimeOperators}).  Making the simplest assumption, we neglect interference between
Wilson coefficients (third term) in
Eq.~(\ref{eq:3bodyGammaZ}). Experimental bound of this simplified
expression now confines $\left(|C_1^{Z'}+\tilde C_1^{Z'}|,
|C_3^{Z'}+\tilde C_3^{Z'}|\right)$ to lie within an ellipse with
semiminor and semimajor axes as upper limits
\begin{subequations}
  \label{par-Z}
  \begin{align}
    y \left| B_{12}^{d_L}\, B_{13}^{d_R}+
      B_{12}^{d_R}\,  B_{13}^{d_L}\right| &<  2.6\E{-4},\\
    y \left| B_{12}^{d_L}\, B_{13}^{d_L}+ B_{12}^{d_R}\,
      B_{13}^{d_R}\right| &< 7.1\E{-4}.
\end{align}
\end{subequations}
Complementary bounds, involving the same couplings and $y$, originate
 from meson mass splittings and CP violation in kaon system and have
 been derived in Ref.~\cite{Langacker:2000ju}
\begin{subequations}
\begin{align}
&y \left|\mathrm{Re}[(B^{d_{R,L}}_{12})^2]\right|<10^{-8},\\ 
&y \left|\mathrm{Re}[(B^{d_{R,L}}_{13})^2]\right|<6\E{-8},\\
&y \left|\mathrm{Im}[(B^{d_{R,L}}_{12})^2]\right|<8\E{-11}.
\end{align}
\end{subequations}
To combine those bounds with Eqs.~(\ref{par-Z}), one should absorb
dimensionless $y$ into the coupling constants by redefinition
$\tilde{B}^{d_{R,L}}_{12(13)} = \sqrt{y}\, B^{d_{R,L}}_{12(13)}$.
However, when we include the constraints given by Eqs.~(\ref{par-Z}), we obtain no further improvement of the bounds on individual $\tilde{B}$ couplings.

Nevertheless, the bounds (\ref{par-R}) and (\ref{par-Z}) are
 interesting since they offer an independent way of constraining the
 particular combination of the parameters, which are not constrained
 by the $B^0_d - \bar B^0_d$, $B^0_s - \bar B^0_s$, $K^0 - \bar K^0$
 oscillations, or by $B^- \to K^- K^- \pi^+$ decay rate (see
 e.g.~\cite{Silvestrini:2005zb}).

Using these inputs we predict the branching ratios for the various
possible two-body decay modes and the $B^- \to \pi^- D^- D_s^+$ decay.
Applying bound (\ref{par-R}) to the RPV model decay rates is
straightforward, except for the $B^- \to \pi^- K^0$ and $B^- \to
\rho^- K^{*0}$ decay modes.  In order to make predictions for these
two modes, we assume as in~\cite{Huitu:1998pa,Fajfer:2000ny}, that
interference term $C_4^{RPV} \tilde{C}_4^{RPV*}$ is negligible, which
leads to the approximation $|C_4^{RPV}-\tilde{C}_4^{RPV}| \simeq
|C_4^{RPV}+\tilde{C}_4^{RPV}|$.

In the case of the $Z'$ model, there are contributions from Wilson
coefficients ``1''~($C_1^{Z'}$, $\tilde{C}_1^{Z'}$) and
``3''~($C_3^{Z'}$, $\tilde{C}_3^{Z'}$). We have already neglected the
interference terms between ``1'' and ``3'' in
Eq.~(\ref{eq:3bodyGammaZ}) to obtain bounds~(\ref{par-Z}) and we
assume that these terms are small for all considered decay modes. Using
Eqs.~(\ref{par-Z}) we can now predict branching ratios for decay modes
$B^- \to \pi^- D^- D_s^+$, $B^- \to \rho^- K^0$, and $B^- \to \pi^-
K^{*0}$.  The remaining two decay rates $B^- \to \pi^- K^0$ and $B^- \to
\rho^- K^{*0}$ can be considered after we neglect interference terms
$C_1^{Z'} \tilde{C}_1^{Z'*}$ and $C_3^{Z'} \tilde{C}_3^{Z'*}$. The
results are summarized in Table~\ref{tab1}.

\begin{table}[!h]
\begin{tabular}{*{121}{l@{\hspace{1.8mm}}}l}
  Decay & SM & MSSM & RPV & $Z'$ \\\hline
  $B^- \to \pi^- \pi^- K^+$ & $3\E{-16}$ & $1\E{-14}$ & $-$ & $-$\\
  $B^- \to \pi^- D^- D_s^+$ & $1\E{-21}$ & $6\E{-20}$ & $2\E{-8}$ & $3\E{-9}$\\
  $B^- \to \pi^- K^0$ & $6\E{-17}$ & $3\E{-15}$ & $1\E{-7}$ & $5\E{-7}$\\
  $B^- \to \rho^- K^0$ & $6\E{-17}$ & $3\E{-15}$ & $1\E{-7}$ & $8\E{-7}$\\
  $B^- \to \pi^- K^{*0}$ & $1\E{-16}$ & $5\E{-15}$ & $6\E{-9}$ & $1\E{-6}$\\
  $B^- \to \rho^- K^{*0}$ & $1\E{-16}$ & $6\E{-15}$ & $7\E{-9}$ & $1\E{-6}$
\end{tabular}
\caption{ The branching ratios for the $\Delta S= -1$ decays of the
  $B^-$ meson calculated within SM, MSSM and RPV models.  The
  experimental upper bound for the $BR(B^- \to \pi^- \pi^- K^+)$ $<
  1.8 \times 10^{-6}$ has been used as an input parameter to fix the
  unknown combinations of the RPV terms (IV column) and the model with
  an additional $Z'$ boson (V column).
}
\label{tab1}
\end{table}
The SM gives negligible contributions. The MSSM is increasing them by
two orders of magnitude, which is still insufficient for the current
and foreseen experimental searches. Using constraints for the
particular combination of the RPV parameters present in the $B^- \to
\pi^- \pi^- K^+$ decay we obtain the largest possible branching ratios
for the two-body decays of $B^- \to \rho^- K^0$ and $B^- \to \pi^-
K^0$, while for the $B^- \to \pi^- K^{*0}$ and $B^- \to \rho^- K^{*0}$
the RPV contribution is suppressed by the renormalization group running.
Their order of magnitude is $10^{-9}$ and thus still experimentally
unreachable. However, these two decay channels are most likely to be
observed in the model with an additional $Z'$ boson, if we assume that
interference terms are negligible.

Since in the experimental measurements only $K_S$ or $K_L$ are
detected and not $K^0$ or $\bar K^0$, it might be difficult to observe
new physics in the $B^- \to \pi^- K^0$ decay mode. Namely, the
branching ratio $BR(B^- \to \pi^- K_S ) = (12.1 \pm 0.7) \times
10^{-6}$ \cite{:2006bi} is two orders of magnitude higher than our
upper bound for the $BR(B^- \to \pi^- K^0)$ making the extraction of
new physics from this decay mode almost impossible.  Therefore, the
two-body decay modes with $K^{*0}$ in the final state seem to be
better candidates for the experimental searches of new physics in the
$b \to d d \bar s$ transitions.

\begin{acknowledgments}
  We are very grateful to J. Zupan for his valuable comments and
  suggestions. This work is supported in part by the Slovenian Research Agency.
\end{acknowledgments}

\appendix

\section{Form factors}
We use the standard form factor parameterization of hadronic current
matrix elements~\cite{Bauer:1987bm, Wirbel:1985ji} between two
pseudoscalar mesons
\begin{eqnarray}
  &&\Braket{ P_2 (p_2) | \bar q_j \gamma^{\mu} q_i | P_1 (p_1) }\nonumber\\*
  &&\quad= F_1(q^2) \left( (p_1+p_2)^{\mu} - \frac{m_{P_1}^2-m_{P_2}^2}{q^2}q^{\mu} \right) \nonumber\\*
  &&\quad\quad+ F_0(q^2) \frac{m_{P_1}^2-m_{P_2}^2}{q^2}q^{\mu}\label{eq:PtoP},
\end{eqnarray}
where $q^{\mu} = (p_1-p_2)^{\mu}$.  We also use the standard decay
constants of the pseudoscalar $K$ and vector $K^{*0}$ mesons
\begin{subequations}\label{eq:decayConstants}
\begin{align} 
  \Braket{ K^0 (p) | \bar d \gamma^\mu \gamma^5 s  | 0 } &= i f_K p^{\mu},\label{eq:KdecayConstant}\\
  \Braket{ K^{*0} (\epsilon_K,p) | \bar d \gamma^\mu \gamma^5 s | 0 }
  &= g_{K^*} \epsilon_K^{\mu*}\label{eq:KstardecayConstant}.
\end{align}
\end{subequations}
Matrix element
between a pseudoscalar and a vector meson is decomposed as customary
\begin{subequations}\label{eq:PtoV}
\begin{align}
  &\Braket{V (\epsilon_V,p_2) | \bar q_j \gamma^{\mu} q_i | P (p_1) }= \frac{2 V(q^2)}{m_P + m_V} \epsilon^{\mu\nu\alpha\beta} \epsilon_{V\nu}^* p_{1\alpha} p_{2\beta},\\
  &\Braket{V (\epsilon_V,p_2) | \bar q_j \gamma^{\mu} \gamma^5 q_i | P (p_1) }=i \epsilon_V^* \cdot q \frac{2m_V}{q^2} q^{\mu} A_0(q^2)\nonumber\\
  &\quad\phantom{=}+i(m_P + m_V) \left[\epsilon_V^{*\mu} - \frac{\epsilon_V^{*} \cdot q}{q^2} q^{\mu}\right] A_1(q^2) \nonumber\\
  &\quad\phantom{=}-i \frac{\epsilon_V^{*}\cdot q}{(m_P +
    m_V)}\left[(p_1+p_2)^{\mu} - \frac{m_P^2-m_V^2}{q^2}
    q^{\mu}\right] A_2(q^2).
\end{align}
\end{subequations}

For the $B^-\to \pi^-$ and $B^- \to \rho^-$ transitions we use form
factors calculated in the relativistic constituent quark model with
numerical input from lattice QCD at high $q^2$~\cite{Melikhov:2000yu}
\begin{widetext}
\begin{subequations}
\begin{align}
  F_1^{\pi B}(q^2) &= \frac{F^{\pi
      B}_1(0)}{(1-q^2/m_{B^*}^2)[1-\sigma_1 q^2/m_{B^*}^2]},
  \quad F^{\pi B}_1(0) = 0.29,\quad \sigma_1 = 0.48,\\
  F_0^{\pi B}(q^2) &= \frac{F^{\pi B}_0(0)}{1-\sigma_1
    q^2/m_{B^*}^2+\sigma_2 q^4/m_{B^*}^4}, \quad F^{\pi
    B}_0(0) = 0.29,\quad \sigma_1 = 0.76, \quad\sigma_2=0.28,
\end{align}
\end{subequations}
\begin{subequations}\label{eq:vectorFormFactors}
\begin{align}
  V^{\rho B}(q^2) &= \frac{V^{\rho B}(0)}{(1-q^2/m_{B^*}^2)[1-\sigma_1 q^2/m_{B^*}^2]}, \quad V^{\rho B}(0) = 0.31,\quad \sigma_1 = 0.59,\\
  A^{\rho B}_0(q^2) &= \frac{A^{\rho B}_0(0)}{(1-q^2/m_B^2)[1-\sigma_1
    q^2/m_B^2]}, \quad A^{\rho B}_0(0) = 0.30, \quad \sigma_1 = 0.54,\\
  A^{\rho B}_1(q^2) &= \frac{A^{\rho B}_1(0)}{1-\sigma_1
    q^2/m_{B^*}^2+\sigma_2 q^4/m_{B^*}^4}, \quad A^{\rho B}_1(0) =
  0.26, \quad \sigma_1=0.73,
  \quad \sigma_2=0.10,\\
  A^{\rho B}_2(q^2) &= \frac{A^{\rho B}_2(0)}{1-\sigma_1
    q^2/m_{B^*}^2+\sigma_2 q^4/m_{B^*}^4}, \quad A^{\rho B}_2(0) =
  0.24, \quad \sigma_1=1.40, \quad \sigma_2=0.50.
\end{align}
\end{subequations}
The transition form factors between heavy mesons $D_s^- \to D^-$ have
been calculated in the chiral Lagrangian approach by the authors in
Ref.~\cite{Fajfer:2004fx}
\begin{subequations}
\begin{align}
  F^{D D_s}_1(q^2)&=0,\\
  F^{D D_s}_0(q^2)&= \frac{q^2}{m_{D_s}^2-m_D^2}\frac{(g_\pi/4)
    f_{K(1430)} \sqrt{m_{D_s}m_D}}{q^2-m_{K(1430)}^2+i\sqrt{q^2}
    \Gamma_{K(1430)}}.
\end{align}
\end{subequations}
The same method has been used to obtain the light to light $K^- \to
\pi^-$ meson transition form factors in Ref.~\cite{Fajfer:1999hh}
\begin{subequations}
\begin{align}
F_1^{\pi K}(q^2) &=\frac{2 g_{VK(892)} g_{K^*}}{q^2-m_{K(892)}^2+i \sqrt{q^2} \Gamma_{K(892)}(q^2)},\\
F_0^{\pi K}(q^2) &= \frac{2 g_{VK(892)} g_{K^*} (1-q^2/m_{K(892)}^2)}
{q^2-m_{K(892)}^2+i \sqrt{q^2}\Gamma_{K(892)}(q^2)}+\frac{q^2}{m_K^2-m_\pi^2} 
\frac{f_{K(1430)} g_{SK(1430)}}{q^2-m_{K(1430)}^2+i \sqrt{q^2}
\Gamma_{K(1430)}(q^2)}.
\end{align}
\end{subequations}
Here the decay widths of the resonances $K^*(892)$ and $K(1430)$ are
taken to be energy dependent~\cite{Fajfer:1999hh}
\begin{subequations}
\begin{align}
\Gamma_{K(892)}(q^2) &= \left(\frac{m_{K(892)}^2}{q^2}\right)^{5/2} 
\left(\frac{[q^2-(m_K+m_\pi)^2][q^2-(m_K-m_\pi)^2]}{[m_{K(892)}^2-(m_K+m_\pi)^2][m_{K(892)}^2-(m_K-m_\pi)^2]}
\right)^{3/2} \Gamma_{K(892)},\\
\Gamma_{K(1430)}(q^2)& = \left(\frac{m_{K(1430)}^2}{q^2}\right)^{3/2}
\left(\frac{[q^2-(m_K+m_\pi)^2][q^2-(m_K-m_\pi)^2]}{[m_{K(1430)}^2-(m_K+m_\pi)^2][m_{K(1430)}^2-(m_K-m_\pi)^2]}
\right)^{1/2} \Gamma_{K(1430)}.
\end{align}
\end{subequations}

\section{Numerical parameters}
\begin{table}[!h]
\centering\begin{tabular}{*{121}{c@{\hspace{6.3mm}}}c}
\hline\hline
$m_B$ & $m_{B^*}$ & $m_\pi$ & $m_\rho$ & $m_{K^+}$ & $m_{K^0}$ & $m_{K^{*0}(892)}$ & $m_D$ & $m_{D_s}$ & $m_{K(1430)}$ & $m_b$ & $m_s$ & $m_d$\\
5.27 & 5.32 & 0.140 & 0.77 &0.494 & 0.498 & 0.892 & 1.87 & 1.96 & 1.41 & 4.2 & 0.10 & 0.006\\
\hline\hline
\end{tabular}
\caption{Meson and quark masses (in GeV) used in our calculations are taken from PDG~\cite{Eidelman:2004wy}.}
\end{table}
Decay constants of the pseudoscalar $K^0$ and vector $K^{*0}$ mesons
are $f_K = 0.160\e{GeV}$~\cite{Melikhov:2000yu} and $g_{K^*} =
0.196\e{GeV}^2$~\cite{Fajfer:2004fx}, respectively.  Further numerical
parameters relevant for the $D_s^- \to D^-$ and $K^- \to \pi^-$
transitions are~\cite{Fajfer:2004fx,Eidelman:2004wy} $g_\pi = 3.73$,
$f_{K(1430)} = 0.05\e{GeV}$, $\Gamma_{K(1430)} = 0.29\e{GeV}$,
$g_{SK(1430)} = 3.7\e{GeV}$, $g_{VK(892)} = 4.59$, and
$\Gamma_{K(892)} = 0.051\e{GeV}$.
\end{widetext}
\bibliography{literatura}

\end{document}